**Title:** Using the Life Grid Interview Technique in Science Education Research

**Authors:** Ashley A. Rowland[1], Dimitri R. Dounas-Frazer[2,3,4], Laura Ríos[2,3], H. J. Lewandowski[2,3], Lisa A. Corwin[1]

**Affiliations:** [1] Department of Ecology and Evolutionary Biology, University of Colorado, Boulder, Boulder CO 80309; [2] Department of Physics, University of Colorado, Boulder, Boulder CO 80309; [3] JILA, National Institute of Standards and Technology and University of Colorado Boulder, Boulder CO 80309; [4] Department of Physics and Astronomy, Western Washington University, Bellingham WA 98225

**Corresponding author:** Lisa A Corwin, Department of Ecology and Evolutionary Biology, 1900 Pleasant St., 334 UCB, Boulder CO 80309

**Abstract**

**Background**: Qualitative interviewing is a common tool that has been utilized by Science, Technology, Engineering, and Mathematics (STEM) education researchers to explore and describe the experiences of students, educators, or other educational stakeholders. Some interviewing techniques use co-creation of an artifact, such as a personal timeline, as a unique way to elicit a detailed narrative from a respondent. The purpose of this commentary is to describe an interview artifact called a life grid. First used and validated in medical sociology to conduct life course research, we adapted the life grid for use in research on undergraduate STEM education. We applied the life grid interview technique to two contexts: 1) students in an advance degree program reflecting on their entire undergraduate career as a biology major, and 2) students in an undergraduate physics program reflecting on a multi-week lab project.

**Results**: We found that the life grid supported four important attributes of an interview: facilitation of the respondents' agency, establishment of rapport between interviewers and respondents, enhanced depth of the respondents' narratives, and the construction of more accurate accounts of events. We situate our experiences with respect to those attributes and compare them with the experiences detailed in literature.

**Conclusions:** We conclude with recommendations for future use of the life grid technique in undergraduate STEM education research. Overall, we find the life grid to be a valuable tool to use when conducting interviews about phenomena with a chronological component.

# Introduction

Qualitative research methodologies, when applied to education, can elucidate educational mechanisms and provide detailed insights into how and why learning happens through rich and nuanced data collection. Qualitative approaches also allow educators to develop a better understanding of both student and educator experiences as well as examine the effects of institutional structure on education. Numerous qualitative research studies have led to breakthroughs in Science, Technology, Engineering, and Math(STEM) education; however, in some STEM education sub-fields, qualitative methodologies are underrepresented compared to quantitative techniques (e.g., in biology education research, Lo et al., 2019). This trend may be due to the recent maturation of a discipline-based STEM education field compared to those with a longer history of conducting discipline-based education research like chemistry and physics (Arthurs, 2019; Lo et al., 2019). Additionally, STEM education researchers whom are first trained in a science discipline may be less likely to employ qualitative techniques and instead rely on the quantitative methodologies most familiar to them (Lo et al., 2019). Thus, those conducting discipline-based education research from emerging fields or trained primarily to use quantitative techniques could greatly benefit from understanding and utilizing qualitative techniques. Likewise, fields with longer histories of qualitative research can benefit from exploring new qualitative techniques or adaptations of existing techniques that emerge in other disciplines in order to enrich their own qualitative methodologies (cf. Robertson et al., 2018). This commentary looks to the field of medical sociology to describe a beneficial qualitative interview technique that can be adopted and adapted to benefit STEM education research: *the life grid interview*.

**Engineering collaborative interviews using artifacts**

Interviewing is a common qualitative technique that, despite being time and resource intensive, is particularly suited for exploring individuals' experiences. Interviews can provide detailed descriptions, characterize processes and mechanisms, and elucidate how events are interpreted by the interview respondent (Weiss, 1994). These in-depth descriptions make it possible for the researcher to produce a report that allows readers to put themselves in the respondent's shoes and develop a deeper understanding of the respondent's experience (Weiss, 1994).

Interviewing is a challenging skill that requires researchers to consider how their interview methodology will ensure that the portrayal of the phenomena under investigation accurately represents an individual's experiences. If proper considerations are not made, recall bias may be introduced, poor rapport can prevent development of detailed narratives, and an interviewer's own personal background and objectives can influence the "restorying" of a respondent's narrative. One way to mitigate these issues is to design the interview to promote collaboration between the interviewer and the respondent, enabling the respondent to assert ownership over their narrative (Creswell & Poth, 2016). A respondent is made to feel empowered when the interview is emphasized as a shared task between the interviewer and the respondent (Gubrium & Holstein, 2001). Thus, in order to best achieve research goals, it is often worth considering how modifications to more standard interview techniques might support collaboration.

The use of artifacts during a narrative interview can serve to promote collaboration, help elicit detail, and direct the conversation. Artifacts can be printed materials, photos, memorabilia, or other objects that serve as a visual cue or talking point discussed during an interview (Creswell & Poth, 2016). Using artifacts during interviews can give respondents agency in the research process. For example, a card sorting exercise is an artifact-based technique that enables respondents to guide and interpret a conversation. In a study of engineers' identities, student engineers were provided 126 cards inscribed with terms corresponding to different engineering identities (Tonso, 2006). Students were asked to sort the most frequently elicited terms into "categories that make sense to you" and to "tell me why you put terms together in each group and to describe how the categories differ." The exercise allowed the researcher to characterize student engineer identities and demonstrate how they arise and are related. Moreover, using artifacts created by respondents themselves can improve the depth and accuracy of interviews. For example, the visual presence of laboratory notebooks while interviewing graduate students was used to explore students' experiences when learning how to effectively use a notebook for scientific documentation (Stanley & Lewandowski, 2016). In another study, researchers conducted structured interviews with physics professors using the professors' own teaching materials to investigate their beliefs and values about teaching and learning (Henderson, Yerushalmi, Kuo, Heller, & Heller, 2007; Yerushalmi, Henderson, Heller, Heller, & Kuo, 2007). In each case, the visual artifacts were used to direct conversation, increase respondent engagement, and prompt richer, more accurate narratives. Yet, despite these examples, descriptions of artifacts used to facilitate qualitative interviews in STEM education contexts is uncommon. Therefore, discipline-based education researchers interested in exploring new qualitative methods may benefit from expanding their repertoire of artifact-based interview techniques.

**Purpose**

In this commentary, two discipline-based education research groups examining biology and physics educational contexts describe two unique uses for an interview artifact, a *life grid*, to facilitate collaboration in an interview. We identified this technique from the field of medical sociology, where it was first piloted in health research and shown to reduce recall bias (Blane, 1996; L. R. Berney & Blane, 1997). We first describe the structure and initial development of a life grid, then elaborate on the unique application of life grid interviews in biology and physics education research. To our knowledge, these applications are the first examples of life grids employed in STEM education research. Drawing on our experiences in STEM and those of others in diverse fields, we describe how this interview technique facilitates important attributes of interviewing, including promoting agency of the respondent, building interviewer-respondent rapport, and enhancing the depth and accuracy of recall. Our aims are to translate the utility of the life grid for STEM education research and encourage researchers to consider employing various qualitative techniques in their studies.

## What is a life grid?

A life grid is a visual grid (i.e., spreadsheet) that is an artifact used during the interview process. There are three key features of every life grid:

1. The entries in the grid represent dimensions of the respondent's life experiences over a specific period of time.
2. The grid row headings are discrete periods of time and the column headings are dimensions of the respondent's life. All headings are determined by the researcher in accordance with the research questions.
3. The majority of grid spaces are left intentionally left blank, to be filled out collaboratively by the interviewer and respondent during the interview.

Abbreviated examples of the life grid interview tool are shown in Tables 1 and 2, while the full life grid examples can be found in Supplemental Tables 1 and 2.

A researcher must consider the context of the study and research questions when selecting which time periods and life dimensions to include in the grid. The life dimension columns are divided into indirect and direct indicators. Indirect indicators are typically highly memorable events that are more factual in nature and not directly related to the research questions. They primarily serve to orient the respondent to the time frame and context of the study but can sometimes lead to discussion of topics that are relevant to the research questions. They can be personal for the respondent, such as place of residence or activities that they participated in, or external to the respondent's life, such as historical, cultural, or social events. Grid entries for indirect indicators may be partially filled in by the researcher in advance of the interview using other ancillary data collected from the project, such as a reflection prompt or transcript. Direct indicators, on the other hand, serve to generate rich discussions around different dimensions of the respondent's life. They are selected carefully to guide the interview dialogue to topics central to addressing the research questions. The direct indicators refer to events that occurred over the same period of time as the indirect indicators.

The life grid, as described above, was first piloted by Blane (1996) to identify the causal and non-causal variables of chronic obstructive airway disease (Blane, 1996). Berney and Blane (1997) next tested the life grid technique to see if it could reduce recall bias compared to traditional interview techniques. They used archived material of the respondents' social circumstances recorded 50 years previously and identified archive items that could be included as indicators on the life grid. They then interviewed respondents about social circumstances during their youth and childhood and compared their account to the archive material. They found that a substantial majority of subjects recalled simple socio-demographic information after a period of 50 years with accuracy (L. R. Berney & Blane, 1997). Parry and colleagues (1999) set out to illustrate the more indeterminate aspects of life grid interviews in a study of smoking behavior among elderly respondents who have a smoking related illness. They described how the life grid initiated discussion around topics relevant to the study, established researcher-respondent rapport, made it easier to return to important topics, confirmed the accuracy of dates, allowed respondents to assert influence over their own biographical accounts, and encouraged the respondent to discover associations between different events in their lives that they had not previously considered (Parry et al., 1999)

Numerous studies have since described benefits of life grid use for life course research. However, we are aware of only a single study, to date, that used the life grid to conduct education research. Abbas, Ashwin, and McLean (2013) conducted life grid interviews to explore the relationship between university rank (i.e., quality in teaching/learning) and inequality between students. They found that the life grid enabled easier comparative checks of data collected at multiple points in a longitudinal study, and they felt that they built a relationship with the respondents, gaining useful knowledge of respondents' lives (Abbas et al., 2013). In the next section, we extend prior explorations of the life grid research with descriptions of unique applications of the life grid in two undergraduate science education contexts.

**Using the life grid technique in two different STEM contexts**

**Study contexts and respondents**

The biology education team (authors A.A.R and L.A.C.) and the physics education team (authors D.R.D, L.R., and H.J.L) collaborating here conducted discipline-based education research in their respective departments.

The biology team used semi-structured interviews to investigate how disciplinary interests and career goals influence undergraduates' pursuit of the critical experiences required to pursue advanced biology degree paths. "Critical experiences" are activities or interactions that students engage in during their undergraduate tenure that are not part of normal curricula and are valued by advanced degree program gatekeepers (e.g., medical school admissions committees). The biology team aimed to construct a timeline of each student's critical experience pursuits during their undergraduate tenure and capture rich narratives that described how and why students accessed such experiences. Respondents in the biology study consisted of former undergraduate students that majored in biology at one of two large, public, PhD-granting institutions in US and had been accepted to either medical school or graduate school for biology. Respondents were asked to reflect on their undergraduate career, and thus were sometimes asked to recall periods up to 10 years in the past.

The physics team conducted semi-structured interviews to investigate how teaching practices can foster or inhibit students' engagement in experimental modeling and their development of a sense of project ownership. "Experimental modeling" refers to the recursive process through which scientists construct and revise models and apparatus in order to achieve agreement between predictions and data (Dounas-Frazer & Lewandowski, 2018). "Project ownership" refers to a student's feelings of agency, responsibility, investment, and pride with respect to their project (Dounas-Frazer, Stanley, & Lewandowski, 2017). They aimed to reconstruct timelines of student participation in a several weeks-long project to reveal connections between modeling and ownership. Respondents in the physics study consisted of undergraduate students currently enrolled in physics programs at doctoral, master's, and baccalaureate colleges in US. The students were enrolled in upper-division lab courses that incorporated multi-week projects. The interviews were conducted immediately after the time interval of interest (i.e., during the last portion of the respondent's final project). Thus, the longest period that respondents were asked to recall was the latter half of a semester, seven weeks prior to the interview.

**Life grid design**

The biology team designed a grid with rows corresponding to each semester of each year during respondents' undergraduate tenure. The life grid also included rows for pre-undergraduate and post-undergraduate years enabling the respondent to mention significant events for these periods. The four indirect indicator columns included large moves/changes in residence, family events and relationships, non-academic events and activities, and financial support (Supplemental Table 1). The six direct indicators columns related to the research questions (e.g., interests and career goals) and reflected the variety of critical experiences desired by advanced degree programs (Association of American Medical Colleges, n.d.-b, n.d.-a). They included academic interests, career goals, participation in research activities, interactions with STEM faculty or advisors, internships/volunteer positions/work related to academics, and participation in professional societies, organizations, or academic living-learning communities. A truncated version of the biology life grid is shown in Table 1, and the full version is available as Supplemental Table 1.

**Table 1. Abbreviated Biology Life Grid**

|  | **Residence** | **Family events and Relationships** | **Interests and Career Goals** | **Activities (research, volunteering, clubs, etc.)** |
|---|---|---|---|---|
| **Year 1 Fall** | *Lived in Dorms* | *Ended relationship from high school.* | *Declared a biology major. Wanted to be a doctor.* | |
| **Year 1 Spring …** | *Lived in dorms with new roommate* | *Grandmother passed away.* | *Took immunology and loved it.* | *Joined the Pre-Health club. Volunteered for the Health Fair.* |

**Legend.** Column 1 shows times that are relevant to the study. Columns 2 and 3 represent indirect indicators. Columns 4 and 5 represent direct indicators. Fictitious representative responses are italicized. Blank spaces exist and are expected in the biology life grid application.

The physics team designed a grid with rows corresponding to weeks spent working on the project. The indirect indicators (column 1) corresponded to dimensions of respondent experiences on the project and included the details of respondents' technical progress on their project. The direct indicators focused on modeling of experimental systems, including respondents' revisions to apparatus or models and changes in overall experimental project goals or sub-goals, as well as ownership, including memorable moments, personal contributions, and experiences working with other students and professors during the project. A truncated version of the physics life grid is shown in Table 2, and the full version is available in the supplemental material (Supplemental Table 2).

**Table 2. Abbreviated Physics Life Grid.**

|  | **Experimental progress** | **Changes to experiment** | **Personal contributions** |
|---|---|---|---|
| **Week** | What were you working on each week? | Did you make changes to your equipment or theories? | What ideas, skills, or work did you personally contribute? |
| *1* | *Assigned to team/project; Interested in design aspects* | | |
| *2* | "We went through all the projects and discussed them. I picked this because it was the most engineering related." | "Drew some initial designs, and then made changes as necessary when designing in software." | "We all contributed pretty equally." |

**Legend.** Week 1 represents what students see pre-interview and Week 2 represents what the grid might look like post interview. Column 1 shows times that are relevant to the study. Column 2 represents the indirect indicator. Columns 3 and 4 represent direct indicators. Examples of actual responses shown in quotes. Blank spaces exist and are expected in the physics application.

**Life grid interview implementation**

Both the biology and physics teams conducted the majority of interviews online using videoconferencing software. Screen sharing was used to ensure that the respondent could view the life grid as it was filled out electronically by the interviewer. Both teams also paid special attention to the way in which the grid was introduced to the respondent. The rows and columns were described and the process of filling it out was framed as a collaborative activity. Before beginning the interview, the biology team reminded the respondent that the interview would be recorded to capture more of the respondent's thoughts than would be possible using only the notes and wrote sparse notes on the grid. Alternatively, the physics team, in addition to recording the interview, wrote continuously throughout the interview to convey to the respondent that all of their descriptions regarding their project were of equal importance.

For the biology study, the interviewer would note each instance when a critical experience was mentioned by the respondent. Whenever this occurred, the interviewer asked a set of follow-up questions to capture the essential details surrounding the respondent's pursuit of, and engagement in, the critical experience including any outcomes of their participation. Similarly, the physics team asked follow-up questions designed to elicit information about respondents' emotional responses that occurred while revising their experiment or experimental goals in order to identify links between modeling and ownership.

Unlike the biology team, which began the interview with a blank life grid, the physics team examined respondents' responses to weekly project surveys and incorporated relevant information on the life grid in advance of the interview. These advance grid entries served as anchor points to aid respondent recall and conveyed a sense of investment by the interviewer. Furthermore, the physics team used these weekly survey responses to generate a more detailed *shadow grid* that only the interviewer could see. The shadow grid was a paper version of the life grid that the interviewer used to prepare for the interview and to jog the respondent's memory or ask follow-up questions during the interview.

Notably, both the biology and physics implementations satisfied the three key features of a life grid: they each used a visual grid comprising entries that represent dimensions of the respondent's life experiences over a specific period of time; row (time intervals) and column (life dimensions) labels were determined by the interviewer in accordance with research questions; and grid entries were filled in during the interview via a collaborative process that involved discourse between the interviewer and respondent. These resulted in the identification of common attributes that enhanced both groups' research (see *Attributes of the life grid technique*). Nevertheless, differences in research questions, temporal scope, and access to background information for respondents resulted in the creation of two unique life grid structures and implementations. Furthermore, each team experienced slightly different procedural outcomes of the interviews resulting from the unique contexts to which the life grid was applied and the slight differences in implementation. These details and differences are described in Supplement 1 for those wishing to read a more detailed methodology of the two life grid implementations.

## Attributes of the life grid technique

Drawing from the existing literature and our personal experiences, we describe four attributes of the life grid technique that make it a unique and effective interviewing tool for studies of undergraduate science education. With thoughtful framing of the interview, we found that the life grid technique can a) help to facilitate respondents' agency, b) establish rapport between interviewer and respondent, c) affect the depth of the respondent's narrative, and d) construct a more accurate account of events.

These attributes are a result of the relationship between the interviewer, respondent, and the life grid as both an artifact and task. Below, we elaborate on the importance of each attribute and draw on relevant literature as well as our own experiences conducting STEM education research to demonstrate how the life grid addresses each attribute.

### Facilitating respondents' agency

*Agency* refers to the capacity, condition, or state of exerting power over a situation, process, or object. Modern interviews seek to secure the constructive voices of the respondents by providing respondents agency to direct construction of their narrative in concert with the interviewer. Gubrium and Holstein (2001) argue that interviews should not be conducted as a discreet neutral set of question and ensuing responses, and instead should be co-constructed and informed by the cultures and experiences of both the interviewer and respondent. Because the interviewer is inherently more agentic at the start of an interview, the respondent's agency should be privileged and supported (Gubrium & Holstein, 2001). This interview philosophy is supported by use of the life grid technique.

Compared to a classic semi-structured interview approach, where the interviewer directs the flow of conversation using questions and follow-ups, the life grid approach allows more choice and direction to come from the respondent. As an artifact, the grid's visual presence and use in questioning allows the respondent to question, contradict, or re-direct the conversation because they can see and make reference to what the interviewer is writing. Prior life grid researchers have described this as "productive interference" during the interview (e.g., Nico, 2016), which increases respondents' agency by allowing them to exert power over and redirect their narrative (Harrison et al., 2011). They argue that this, in turn, enhances the relevance of information (Parry et al., 1999; Sheridan, Chamberlain, & Dupuis, 2011). In addition, they describe how the collaborative task of constructing the life grid puts the respondent on equal footing with the interviewer, allowing more agentic moves by the respondent (Groenewald & Bhana, 2015; Parry et al., 1999; Wilson et al., 2007). Since the life grid allows respondents to exert a high degree of control over the pace, structure, and emphasis of the interview, respondents can draw more attention to personally salient events (Parry et al., 1999), avoid topics that cause them discomfort or stress (Crawford and Wilkinson, 2018; Richardson et al., 2009), or discuss difficult events by focusing attention on the shared task of filling in the grid (Crawford and Wilkinson, 2018).

Like prior life grid users, we felt that the collaborative completion of the life grid affords respondents the power and opportunity to guide the discussion. Specifically, we noted agentic instances when respondents a) directed the interview in an unanticipated way and b) guided the interview to discuss relevant topics that shaped the respondent's experiences but were not directly related to grid columns or interviewer questions. Often, the details added by the respondent were temporally linked to the life grid and served to form a more complete image of the individual's narrative. The following example from the biology application illustrates how Sarah (all respondent names are pseudonyms) interacted with the life grid to bring up a relevant, somewhat sensitive, experience that was not directly related the interview indicators under discussion. This occurred after the interviewer asked Sarah if there were other events she wanted to add to a column on the grid.

> Sarah: Oh, one other thing did happen, sorry, I'm sure it happens a lot but, I got an MIP (minor in possession), like, right when I got to college.
>
> Ashley: Oh, your freshman year?
>
> Sarah: There's no column for it (laughing)
>
> Ashley: Well, it's a non-academic event, we'll say. (laughing)
>
> Sarah: "interactions with the law" (laughing). Yeah, I got an MIP and after that I was really nervous, but it got sealed.
>
> Sarah: [describes the incident] …
>
> Sarah: I always was nervous that [the MIP] might surface and thought that I might not be able to get into medical school if I have an MIP.
>
> Ashley: Okay, was that part of the reasons you said you were considering like nursing and other trajectories you were interested in?
>
> Sarah: Yeah and I think that's one of the reasons that I did so well in school, 'cause I didn't give myself wiggle room with my GPA to compensate for that.

Sarah's agency became apparent as she utilized the temporal nature of the grid to discuss an event that was important for that time period of her life but was not explicitly represented on the life grid and was unanticipated by the interviewer. Another example from the biology application illustrates how the visual nature of the life grid enabled Miguel to elaborate on his story.

> Ashley: Do you have anything else that you would like to add? [gestures to grid]
>
> Miguel: No, I think that's good.
>
> Ashley: Awesome.
>
> Miguel: [looking at grid] One thing that we really didn't talk about much was how I think, for a lot of people, it's important to be able to cope with things that come up in your pursuit of whatever career you have.
>
> Ashley: Yeah.
>
> Miguel: I think for me, like I talked about this a lot with my postdoc [referencing a cell on grid that refers to research mentor interactions] because we were studying stress and the mechanisms, and I believe my faith was a big way that I was able to sort of detach from the stress and think of it in perspective and learn about things that are, you know, that don't have to do with school and there's a bigger picture.

Miguel's connection with his postdoc, referenced explicitly in a direct indicator cell on the grid, allowed him to elaborate on an aspect he considered important in his narrative.

Despite our personal experiences and published studies describing how the life grid technique facilitates respondents' agency, there are caveats to consider. In a study on adolescent substance abuse, Wilson (2007) described how one respondent reacted to the blank life grid with discouragement, stating that they knew that their complex story would not "fit" within the bounds of the grid. This lack of "fit" may have resulted in this respondent feeling unable to act to sufficiently express his story (i.e., he may have felt a lack of agency). Therefore, we recommend that researchers interested in this technique consider how they will introduce the grid to the respondent and actively facilitate discussion to support respondent agency.

**Establishing rapport between interviewer and respondent**

*Rapport* refers to a positive, friendly relationship or connection between two people most often characterized by general agreement, mutual understanding, or empathy that makes communication fluid and easy. Building positive rapport is a critical component of an effective interview. Kvale and Brinkmann (2009) describe how a skilled interviewer creates a positive relationship that leads to the production of significant knowledge. They, like Gubrium and Holstein (2001), describe an interview as a co-constructive process involving both parties rather than a monologue recorded by an interviewer (Kvale and Brinkmann, 2009).

The collaborative construction of the life grid by the interviewer and respondent often fosters a relationship in which positive feelings arise and the interviewer better understands the feelings and needs of the respondent. In other words, positive rapport is developed (Wilson et al., 2007). Specifically, life grid co-construction facilitates an indirect, less confronting way of eliciting information, which supports rapport building. This is in contrast to direct questions often used in more traditional interview techniques, which can feel combative and make the implicit interviewer-respondent hierarchy more apparent (Groenewald & Bhana, 2015; Harrison et al., 2011; Parry et al., 1999; Wilson et al., 2007). Furthermore, co-construction and completion of the life grid can foster a sense of satisfaction and accomplishment among both the interviewer and respondent (Crawford and Wilkinson, 2018; Richardson et al., 2009). This shared sense of accomplishment may inspire lasting positive relationships and rapport, which can be leveraged over multiple interviews. This benefit is described by Abbas, Ashwin, and McLean (2013), who used initial life grid interviews to help build rapport and facilitate lasting relationships with respondents over three years of interviews.

Our interviews displayed similar patterns as previous literature with regard to rapport. Specifically, we noted instances of positive rapport when respondents expressed positive emotions related to their co-construction of the grid or when expressions of agreement, mutual understanding, or empathy emerged during an interview. In the physics study, the interviewer asked each respondent how it felt to see the completed grid. Some respondents expressed happy surprise: "It's kind of crazy. [...] It feels good!," and, "There's a lot more here than I thought!" Others described the grid as a helpful or logical way to structure their narrative: "It's an interesting way to lay things out. [...] a helpful way to break things down," and, "It's a fairly logical progression." Similar sentiments were shared by the interviewer, and hence implementation of the life grid technique yielded many instances of mutual satisfaction among respondent and interviewer.

Also, like findings by Parry and colleagues (1999), the biology team found that discussing topics of interest to the research became easier after working through the initial indirect indicators, which provided opportunities to develop rapport. For example, indirect indicators often allowed the interviewer to find commonalities with the respondent that were unrelated to the study, allowing the interviewer to express understanding or empathy and build rapport. In the quote below, the biology team interviewer, Ashley, identified a commonality between themselves and respondent Emily during discussion of indirect indicators.

> Emily: And I was still on the [sport] team.
> Ashley: You did [sport name]?
> Emily: Uh-uh (affirmative).
> Ashley: All right, I [did that sport] in college also. Did you do that from your freshman year? I should write that on here [referring to the grid].
> Emily: Yep.
> Ashley: So, you [participated] all four years?

> Emily: Well, I [did it] for the first two, and then I switched to [another sport] for the second two-
>
> Ashley: Oh.
>
> Emily: ... because I [had a bad injury].
>
> Ashley: Oh my God.
>
> Emily: So my coach was like, "Are you really sure that you want to continue with this?"
>
> Ashley: Sounds like some of my friends [that had similar injuries].
>
> Emily: Yeah.

As seen here, working through indirect indicators (*Non-academic events* in the example above) frequently fostered mutual understanding or highlighted shared experiences and facilitated personal exchanges between interviewer and respondent. This is likely in part because the indirect indicators were general enough to allow connections to be made.

In contrast to the biology group, the physics interviewers used reflections to fill out some indirect indicators prior to the interviews, personalizing their grid to the respondent and creating highly specific indirect indicators. They also created the shadow grid to inform follow-up questions during the interview. This may have demonstrated to respondents that the interviewer was invested in their story, contributing to positive rapport. Overall, despite slightly different approaches, the use of indirect indicators facilitated rapport building exchanges for a majority of students in both the physics and biology study. Therefore, careful considerations of which indirect indicators to include or which information to pre-fill in advance is likely to be important in facilitating this function of the life grid. Future work could explore whether and how aspects of each implementation contributed to rapport building between interviewers and respondents.

**Affecting the depth of the respondent's narrative**

Kvale and Brinkmann (2009) enumerate various aspects of qualitative interviews including that they are descriptive and specific. A descriptive interview involves encouraging the respondent to describe as precisely as possible what they experience and feel. The specificity of the interview refers to eliciting detailed descriptions of situations and actions, rather than general opinions, to inform meaning (Kyale and Brinkmann, 2009). We refer to the high level of detail, information, and connections between ideas that a person includes in a verbal recounting of their experiences as *depth,* and find that it is supported by the life grid interview technique.

Several researchers have found that the life grid approach can facilitate collection of rich, qualitative narratives during the interview (Groenewald & Bhana, 2015; Nico, 2016). They cite the development of rapport and facilitation of respondents' agency as two aspects that contribute to this attribute. Specifically, increased agency and the ability to cross-reference events allows surprising connections between the research topics and other, seemingly unrelated, areas of the respondent's life to be made (Harrison et al., 2011; Nico, 2016) and rapport increases respondents' comfort and thus their willingness to share their narrative (Groenewald & Bhana, 2015; Parry et al., 1999). Beyond these aspects, the life grid as an artifact can be used to ask questions in creative ways that elicit more information than traditional interview techniques. For example, instead of asking a respondent a direct question such as "Describe your experiences engaging in undergraduate research." and then following up with "What were your reasons for engaging in these experiences," an interviewer can point to the "Research Experience" column on the grid and simply state "let's begin filling out this column." This focus on the life grid artifact is less confrontational and allows respondents to elaborate on their stories since they are indirectly discussing life experiences via the life grid (Nico, 2016). Considering the grid as an artifact also facilitates greater detail. When sufficient fluidity is allowed during the interview (e.g., movement freely around the grid during discussion), the grid structure can allow respondents to visualize and discuss connections between seemingly unrelated events (Blane, 1996; Parry et al., 1999). Fluidity also facilitates return to prior events when new connections are made (Nico, 2016). Thus, the grid combines the historical occurrence of events with the respondent's own interpretation and subjective experience, enriching and extending the narrative (Nico, 2016).

However, eliciting depth and nuance does not arise solely through incorporation of a life grid in an interview. Specific facilitation techniques enable this process. Nico (2016) specifically cites that flexibility in filling out the grid should be used to facilitate connections. Allowing respondents to move backwards and forwards through time enhances the detail and connections that can be made and allows more recall to occur. Likewise, they describe that the grid should be visible and comprehensible to respondents (e.g., abbreviations should be avoided) during the interview to facilitate respondents' use of the life grid to direct their own narrative. Nico (2016) describes that introduction of the life grid and description of how it should be used is critical to facilitate these features and enhance interview depth. We also found that a thoughtful introduction of the life grid at the start of an interview to be an important step in eliciting depth.

In both biology and physics studies, we introduced the life grid as a tool to help build a narrative, encouraged respondents to backtrack and fill in the grid as appropriate, and paused to ask respondents follow-up questions frequently during the interview. Specific instances of added depth and description occurred when respondents cross-referenced information in different columns of the grid and provided insights into the reasons why they took certain actions, engaged in specific tasks, or felt different emotions. Added depth also emerged during participants' explanations regarding how they had used the grid to build their narrative during the interview. In an example from the biology application, an interaction between the interviewer, Ashley, and the respondent, Megan, illustrates how the visual nature of the grid allowed greater temporal clarity and more detail regarding how living conditions (an indirect indicator) influenced Megan's academic major and advising (related to direct indicators).

> Ashley: So did your advisors help you plan your classes, your schedule? (referring to a direct indicator column on the grid)
>
> Megan: Not really. I had a friend who [did], one of my roommates was incredibly organized and so she sat down [and looked at] my schedule every semester, and just like, "Does this the sound interesting?" I was like, "Yeah, sure, I'll take that."
>
> Ashley: Wow. Is that a friend that you had met in the dorms (pointing to the indirect indicator column on the grid)?
>
> Megan: Yeah. No, she was my roommate sophomore junior (pointing to a different row on the grid). Well actually she lived across the hall from me in my freshman year, and she was the one that figured out that I could technically be an Anthropology major (referring to another direct indicator column on the grid). Then we lived together sophomore junior and senior year. So I think I only had help my senior year from my major advisor (pointing to a different direct indicator column).

After being prompted from Ashley, Megan references the life grid as she describes how her living arrangement with a roommate influenced her chosen course of study. She also clarifies that academic advising did not have a role to play in her early undergraduate choices. These techniques added details and facilitated new connections, while also supporting respondents' agency (see quotes from Miguel and Sarah above).

Prompting reflection on blank grid entries was another mechanism through which the life grid facilitated deep and detailed interviews. Similar to what Nico (2016) describes as challenging respondents to fill 'omissions' within their narrative, the physics interviewers asked respondents whether or not information might belong in empty spaces present at the end of the interview. Most respondents indicated that the spaces should be blank. Several provided rationale for leaving some entries unfilled. Several others added information to their narrative. Consider the following excerpt from a physics interview with Sophie, a respondent who was working on an acoustic levitation project:

> Dimitri: Zooming out, big picture, there are a few blank spots, and I wanted to take some time just to look over them. It's okay if we have blank spots. We don't have to fill out every grid entry. I just wanted to make sure you're okay with leaving these spots blank. Or, if there's something, when looking at them, that you wanted to fill out.
> Sophie: Mhm. I guess that the last "interactions with professors" column, it was fun to show our professors it [the project] working. It was fun in general to show people the project working, because it's the sort of thing that, it's cool to see something floating in the air, it's cool to see something levitating. So, kind of at the end of the day, it was fun to show off. Kind of like affirming to have a professor be like, "Wow, that's cool!" (laughs) [...]
> Dimitri: Anything else?
> Sophie: (5 second pause) Um, I don't think so. I think the "personal contributions" column is, like, pretty empty because my group was working on things together most of the time. So, I'm sure that, like, I and my group members all had personal contributions in any given time we [were] working on it together, but none of them really stands out because they were, like, brainstorming or troubleshooting as a group. So, it wouldn't be like, "Ah, yes my contribution was this."

Sophie used this opportunity to add depth to her narrative in two ways. First, she described an affirming interaction with a professor after her apparatus became functional. Second, she provided rationale and context for leaving multiple entries blank in the column designed to probe students' personal contributions to the project. According to Sophie, blank entries in this column were due to the presence of collaborative problem solving, not the absence of individual effort. Thus, the life grid as an artifact facilitated additions to Sophie's narrative by giving her the opportunity to see and reflect upon blank entries.

In addition to asking follow-up questions, the physics team used the shadow grid to support respondents in the shared task of filling out the grid and enhancing the detail of grid entries. For example, when Carlos was trying to recall interactions with his professor, he described patterns of interaction that spanned his project experience, but he initially could not think of any specific interactions. The interviewer used the shadow grid to jog the respondent's memory:

> Carlos: I'm trying to think if there's a specific example. Um. Yeah. I can't think of a specific example right now, but I liked the combination of [instruction] being generally hands-off but also checking with us so that if we were reaching a problem, the professor would talk to us. [...]
>
> Dimitri: I have a couple examples from your surveys that you had filled out [examining the shadow grid]. So, one, you'd mentioned that the professor was the one who suggested that the heat of the laser might be altering the beam intensity. Do you remember that interaction?
>
> Carlos: Um. Yeah, yeah. I'm trying to think of the (trails off). So yeah, that was, again, when we were sitting down looking at how the different trials were not consistent and we were kind of befuddled by that. Trying to come up with reasons. Most of the reasons we were thinking about were some issue with our, basically everything about the apparatus except the laser. We had just assumed the laser was consistent. And so, then the professor coming in and taking this idea that the laser was inconsistent serious, that helped us realize that that could be the problem.

After Carlos elaborated on a specific interaction with his professor, the interviewer asked a clarifying question to help situate that interaction in the context of the pattern of interactions that Carlos had previously described:

> Dimitri: Was this an example of where you all were trying to solve a problem and the professor came in to check in, or did you reach out to her?
>
> Carlos: We did not reach out to her. I mean, looking in hindsight, we should have. But again, we were so focused on a problem, sometimes it's hard to pick your head up. And so having the professor check in every once in a while, was helpful. And that was a great example of how that works.

Thus, the combination of the shadow grid and clarifying questions helped the interviewer and respondent co-construct a detailed account of student-professor interactions, including both abstract patterns of interaction and "great example[s]".

Overall, we found that using the life grid to facilitate rich descriptions of respondent narratives was perhaps the most important function of the life grid in both the biology and physics studies. However, one study from another researcher reported a contrasting experience. In using the life grid technique, Bell (2005) felt that the linear and event-centered structure of the grid discouraged respondents from providing rich details about their experiences. Bell (2005) asserted that the grid caused respondents to focus only on facts without elaborating on feelings, context, and other relevant details. He found that life grid interview respondents tended to work chronologically from the top-most to the bottom-most row, and that respondents were hesitant to go "backward" to discuss previously covered rows. Thus, in Bell's study, interviews were relatively quick and devoid of detail. Therefore, he recommended that the life grid technique be used only to generate factual data and be avoided for more affective and attitudinal topics (Bell, 2005).

We suspect that the incongruity of Bell's results as compared to ours and other studies is caused by differences in grid structure and facilitation techniques. The grid used in the Bell (2005) study included 50 rows to capture details of long-term marriages dating as far back as 50 years. The large size and temporal scope of this grid may have been prevented respondents from providing detail, and instead, directed focus toward completing the task at hand. We made sure to limit the size of the grid to avoid intimidating respondents or creating time pressure. Bell also selected very broad direct indicators such as "Marriage and Family" and "Relationship issues". In contrast, we used more specific direct indicators to encourage detailed narratives from respondents attempting to recall factual events. Finally, we aimed to provide flexibility and actively encouraged participants to backtrack, cross-reference, and return to blank spaces, all recommendations made by Nico (2016) to elicit more detail. We conclude that the way in which researchers introduce and use the grid is highly likely to influence the depth of their data.

**Construct a more accurate account of events**

*Accuracy* in an interview context refers to the correctness and completeness of a person's description of events. Recall bias is systematic error that arises in reports of past events due to differences in the accuracy or completeness of recollections. Recall bias is problematic in that it can introduce inaccuracies during temporal narratives which, at their worst, may change the interpretation of events' causes and consequences.

A classic and frequently cited benefit of the life grid approach is reduction in recall bias in retrospective research, which was the main motivation for using this approach in early studies (L. R. Berney & Blane, 1997; Blane, 1996). Enhancing recall continues to be a benefit of use of the life grid in more recent studies (Groenwald & Bhana, 2015; Harrison et al., 2011; Porcellato et al., 2016). The primary characteristics of the life grid approach that give rise to this benefit are the presentation of a visual timeline used to place events and the use of indirect indicators. Indirect indicators, which are highly memorable events that occur during events of interest to the study, can reflect major societal events, like wars or natural disasters (e.g., Parry et al., 1999; Porcellato et al., 2016), or personally relevant events, like sports or hobbies (e.g., Wilson et al., 2007). Indirect indicator columns can be filled out by the interviewer prior to the interview using information that the interviewer suspects will be memorable to respondents (e.g., Parry et al., 1999) or using previously known information about the respondent (e.g., childhood health records, see Holland et al., 2000). Alternatively, these columns can be filled out during the interview based on events that are personally relevant to the respondent (e.g., Richardson et al., 2009; Wilson et al., 2007). Deliberate selection of indirect indicators grounds the discourse in events that are relevant and memorable to the respondent and may even uncover surprising relationships between life-events and study-relevant information.

Both the biology and physics team found the grid helpful for enhancing accuracy of recall. To improve recall, the biology team used indirect indicators likely to be relevant to college-age students: semester classes taken, changes in living situations, changes in relationship (romantic or friend) status, and large family events (births, deaths, marriages, etc.). Indirect indicator columns were filled out at the start of the interview and subsequently used as markers to ground study-relevant information. The following excerpt illustrates cross-referencing for temporal accuracy.

> Ashley: That's cool. When did that happen?
> Kristin: I guess that was fall of my senior year. Yeah, yeah, fall of my senior year.
> Ashley: Fall of your senior year. Okay, so that was ... Let's see. So that was after you had finished working [searching the grid], you'd already done the research that time and then you were working for [company name] when you were traveling?

> Kristin: Mm-hmm (affirmative), yeah.

In some cases cross-referencing resulted in correction of previously mentioned information.

> Sarah: Yeah, there are two semesters, it was physiology 1 and physiology 2.
>
> Ashley: Okay, so you had moved into the apartment by then [referencing the grid], you were out of the [other living arrangement]?
>
> Sarah: Yeah, I think they make you finish physics and stuff.
>
> Ashley: Do you think that's around the time of your car accident, then, that you were in your second semester?
>
> Sarah: Uh, you put it under summer, but the it was in the fall.
>
> Ashley: Got it. (corrected the life grid entry)

Earlier, Sarah had listed an experience relevant to the study in the summer. In this quote, she corrects her earlier statement and moves the experience to the fall.

The physics team used respondents' progress on their project as an indirect indicator, and the corresponding column of the life grid was filled out by the interviewer based on respondents' responses to weekly surveys. Thus, each respondent was presented with a unique life grid during their interview, which assisted with recall. The interviewer also often used the shadow grid to guide follow-up questions when filling out columns of indirect and direct indicators, thus facilitating recall while avoiding imposing too rigid a structure at the start of the interview (see the example from Carlos above). When asked to reflect on the completed grid, one respondent, Francisco, drew a connection between the structure of the life grid and his ability to recall events:

> Dimitri: So, just looking over the grid, how does it feel seeing the project laid out this way?
>
> Francisco: Going through it from top left to bottom right and going column by column makes a lot of sense. [...] It jogs the memory as you go, and it's a fairly logical progression of, you know, what happened this week, this week, this week, this week.

In summary, all of the attributes described above contribute to the utility of the life grid technique in facilitating a positive experience for the respondent (agency and rapport) and in enhancing the detail and accuracy of the narrative. However, the mere presence of the life grid does not ensure these attributes. We found that how the life grid is structured, introduced, and used during interviews is paramount in realizing the benefits described above. In addition, respondents' unique identities and comfort levels affect interviews regardless of the interview technique, resulting in variable experiences. Overall, we have found that the life grid technique is highly useful in collecting detailed data and nuanced narratives describing a discrete period of time in a students' academic career.

**Conclusions**

In this paper, we have described the life grid technique, explored two specific contexts in which the life grid approach was applied in STEM discipline-based education research, and detailed attributes of life grid use that may add to its utility as an interview technique. Additionally, this work addresses recent calls in discipline-based education research to go beyond simply demonstrating *that* something works and to additionally characterize *how* it works and explain moderating effects of different educational contexts (Tanner, 2011; Dolan, 2015). These calls recognize the need to draw upon theory and methodologies from other fields (e.g., cognitive science, sociology, etc.), much in the way that we adapted the life grid from the field of medical sociology, in order to advance STEM discipline-based education research.

Our experiences using the life grid in STEM education research combined with examples from the literature, primarily in medical sociology, suggest that this technique is most useful to address questions with a substantial temporal component. These include education questions similar to those presented here, which investigate experiences over the course of a project or academic career, and may extend to questions that ask about time periods before or after specific academic events. Constructs that undergo development or are subject to process are inherently well-suited to studies use of a life grid. For example, development of identity, career clarification, and development of expertise might all be appropriate topics for life grid work. The variety of contexts to which the life grid has been applied exemplifies the plasticity of the technique. It can be used for time intervals that span weeks to years, and grid rows and columns can be modified to meet the needs of specific studies. These aspects may make the grid an attractive option for the study of a variety of temporal phenomena in educational contexts.

Despite the many potential uses of the life grid technique, it is not appropriate in all instances. Research questions that are *not* temporal in nature will not benefit from this technique. For example, a research question asking about how an individual perceives a school policy or the culture of a department would not be answered using a life grid. Also, due to the more open-ended nature of the life grid, which allows respondents to guide much of the direction of the interview, highly specific research questions that aim to target very specific details may not benefit from this technique. Instead, techniques such as structured interviews may be used for these types of questions.

Notably, the precise application of the biology and physics life grids varied in a few ways that demonstrate its flexibility, including the timescale under investigation (i.e., years or weeks), use of prior knowledge of the participant (i.e., student reflections and demographic information), and the structure of the grid (i.e., size and use of a shadow grid). Despite these differences, we feel that several specific components of the life grid structure are consistently useful for facilitating the life grid attributes we describe above. These components include the row headings that designate discrete periods of time, column headings that are direct and indirect indicators relating to dimensions of the respondents' life, and grid spaces left primarily blank to be filled out collaboratively during the interview. Also, we found that thoughtful facilitation of the life grid is important for achieving positive outcomes. While we feel that these defining life grid components contributed to the results we observed, relationships between specific life grid components and attributes have not yet been tested, aside from improved recall. Therefore, we cannot yet say which components are essential for facilitating the different life grid attributes. An important next step in determining the utility of this technique would be to test which components of the life grid are essential for it to function as described.

We hope that the examples and descriptions provided here will generate ideas about how to approach and structure interviews in STEM education research in order to generate both accurate and rich descriptions of participants' experiences. We also hope that this work will generate future research on life grid use and function in STEM education contexts. Based on our experiences, we conclude that life grids are a valuable addition to the toolkit of qualitative STEM discipline-based education research methods.

## List of abbreviations

STEM – Science, Technology, Engineering, and Mathematics

# Declarations

**Availability of data and materials**

The de-identified interview datasets used to inform this work are not currently publicly available because other studies analyzing the data are not yet complete and published, but the data are available from the corresponding author on reasonable request.

**Competing interests**

The authors declare that they have no competing interests.

**Funding**

This material is based upon work supported by grants No. DUE-1712160, No. DUE-1611868, No. DUE-1726045, and No. PHY-1734006. The NSF did not contribute to the study design, data collection and analysis, interpretation of the data, or writing of the manuscript. This manuscript solely represents the views of the authors and does not represent the views of the NSF.

**Authors' contributions**

A.A.R conceived and designed the work, lead manuscript writing and project organization, directed the literature review, performed interviews related to the biology project, and synthesized and interpreted exemplar data.

D.R.D and L.R. contributed to manuscript writing and literature review, performed interviews related to the physics project, and synthesized and interpreted exemplar data.

H.J.L contributed to manuscript writing and literature review and synthesized and interpreted exemplar data.

L.A.C conceived and designed the work, contributed to manuscript writing, project organization, and the literature review, and synthesized and interpreted exemplar data.

The initial draft of the manuscript was written via a collaborative writing process that was organized by A.A.R and L.A.C and in which all coauthors participated roughly equally.

**Acknowledgements**

We are grateful for useful input on this manuscript by Angela Little, Laura Wood, Vashti Sawtelle, and Sarah Maestrales from the Advancing Narratives about Student Experiences through Research (ANSER) group at Michigan State University. This material is based upon work supported by the NSF under Grants No. DUE-1712160, No. DUE-1611868, No. DUE-1726045, and No. PHY-1734006.

http://sru.soc.surrey.ac.uk/SRU57.pdf

Robertson, A. D., Scherr, R. E., & McKagan, S. B. (2013). Paradigms in Physics Education Research. Retrieved from http://arxiv.org/abs/1307.4135

Sheridan, J., Chamberlain, K., & Dupuis, A. (2011). Timelining: Visualizing experience. *Qualitative Research*, *11*(5), 552–569. doi:10.1177/1468794111413235

Stanley, J. T., & Lewandowski, H. J. (2016). Lab notebooks as scientific communication: Investigating development from undergraduate courses to graduate research. *Physical Review Physics Education Research*, *12*(2), 1–11.

Tanner, K. D. (2011). Reconsidering "What Works." CBE—Life Sciences Education, 10(4), 329–333. doi:10.1187/cbe.11-09-0085

Tonso, K. L. (2006). Student engineers and engineer identity: Campus engineer identities as figured world. *Cultural Studies of Science Education*, *1*(2), 273–307. doi:10.1007/s11422-005-9009-2

Weiss, R. S. (1994). *Learning from Strangers: The Art and Method of Qualitative Interview Studies*. New York, NY: Free Press.

Wilson, S., Cunningham-Burley, S., Bancroft, A., Backett-Milburn, K., & Masters, H. (2007). Young people, biographical narratives and the life grid: Young people's accounts of parental substance use. *Qualitative Research*, *7*(1), 135–151. doi:10.1177/1468794107071427

Yerushalmi, E., Henderson, C., Heller, K., Heller, P., & Kuo, V. (2007). Physics faculty beliefs and values about the teaching and learning of problem solving. I. Mapping the common core. *Physical Review Special Topics - Physics Education Research*, *3*(2), 1–31.
38